\input harvmac
\input psfig
\input epsf
\noblackbox
\newcount\figno
 \figno=0
 \def\fig#1#2#3{
\par\begingroup\parindent=0pt\leftskip=1cm\rightskip=1cm\parindent=0pt
 \baselineskip=11pt
 \global\advance\figno by 1
 \midinsert
 \epsfxsize=#3
 \centerline{\epsfbox{#2}}
 \vskip 12pt
 {\bf Fig.\ \the\figno: } #1\par
 \endinsert\endgroup\par
 }
 \def\figlabel#1{\xdef#1{\the\figno}}

\def\encadremath#1{\vbox{\hrule\hbox{\vrule\kern8pt\vbox{\kern8pt
 \hbox{$\displaystyle #1$}\kern8pt}
 \kern8pt\vrule}\hrule}}
 %
 %
 
 %%% Paragraphs

 %%% special math symbols
 \font\cmss=cmss10
 \font\cmsss=cmss10 at 7pt
 \def\rlx{\relax\leavevmode}
 \def\inbar{\vrule height1.5ex width.4pt depth0pt}
 \def\IC{\relax\,\hbox{$\inbar\kern-.3em{\rm C}$}}
 \def\IN{\relax{\rm I\kern-.18em N}}
 \def\IP{\relax{\rm I\kern-.18em P}}

\def\ZZ{\rlx\leavevmode\ifmmode\mathchoice{\hbox{\cmss
Z\kern-.4em Z}}
  {\hbox{\cmss Z\kern-.4em Z}}{\lower.9pt\hbox{\cmsss
Z\kern-.36em Z}}
  {\lower1.2pt\hbox{\cmsss Z\kern-.36em Z}}\else{\cmss
Z\kern-.4em
  Z}\fi}
 %%% misc.
 \def\IZ{\relax\ifmmode\mathchoice
 {\hbox{\cmss Z\kern-.4em Z}}{\hbox{\cmss Z\kern-.4em
Z}}
 {\lower.9pt\hbox{\cmsss Z\kern-.4em Z}}
 {\lower1.2pt\hbox{\cmsss Z\kern-.4em Z}}\else{\cmss
Z\kern-.4em
 Z}\fi}
 %%% misc.
 \def\IZ{\relax\ifmmode\mathchoice
 {\hbox{\cmss Z\kern-.4em Z}}{\hbox{\cmss Z\kern-.4em
Z}}
 {\lower.9pt\hbox{\cmsss Z\kern-.4em Z}}
 {\lower1.2pt\hbox{\cmsss Z\kern-.4em Z}}\else{\cmss
Z\kern-.4em Z}\fi}

 \def\narrowplus{\kern -.04truein + \kern -.03truein}
 \def\narrowminus{- \kern -.04truein}
 \def\narrowminussub{\kern -.02truein - \kern
-.01truein}

 \def\frac#1#2{{#1\over #2}}

 \def\IZ{\relax\ifmmode\mathchoice
 {\hbox{\cmss Z\kern-.4em Z}}{\hbox{\cmss Z\kern-.4em
Z}}
 {\lower.9pt\hbox{\cmsss Z\kern-.4em Z}}
 {\lower1.2pt\hbox{\cmsss Z\kern-.4em Z}}\else{\cmss
Z\kern-.4em Z}\fi}
 \def\IB{\relax{\rm I\kern-.18em B}}
 \def\IC{{\relax\hbox{$\inbar\kern-.3em{\rm C}$}}}
 \def\Ic{{\relax\hbox{$\inbar\kern-.22em{\rm c}$}}}
 \def\ID{\relax{\rm I\kern-.18em D}}
 \def\IE{\relax{\rm I\kern-.18em E}}
 \def\IF{\relax{\rm I\kern-.18em F}}
 \def\IG{\relax\hbox{$\inbar\kern-.3em{\rm G}$}}
 \def\IGa{\relax\hbox{${\rm I}\kern-.18em\Gamma$}}
 \def\IH{\relax{\rm I\kern-.18em H}}
 \def\II{\relax{\rm I\kern-.18em I}}
 \def\IK{\relax{\rm I\kern-.18em K}}
 \def\IP{\relax{\rm I\kern-.18em P}}
 %\def\IX{\relax{\rm X\kern-.01em X}}
 %this doesn't work

 \font\cmss=cmss10 \font\cmsss=cmss10 at 7pt
 \def\IR{\relax{\rm I\kern-.18em R}}

 %

 %
 %       \eqn\label{a+b=c}       gives displayed equation, numbered
 %                               consecutively within sections.
%     \eqnn and \eqna define labels in advance (of eqalign?)
 %
 \def\eqnn#1{\xdef
#1{(\secsym\the\meqno)}\writedef{#1\leftbracket#1}%
 \global\advance\meqno by1\wrlabeL#1}
 \def\eqna#1{\xdef
#1##1{\hbox{$(\secsym\the\meqno##1)$}}

\writedef{#1\numbersign1\leftbracket#1{\numbersign1}}%
 \global\advance\meqno by1\wrlabeL{#1$\{\}$}}
 \def\eqn#1#2{\xdef
#1{(\secsym\the\meqno)}\writedef{#1\leftbracket#1}%
 \global\advance\meqno by1$$#2\eqno#1\eqlabeL#1$$}

 \lref\author{Name}
\lref\berg{E.~Bergshoeff,
 D.~S.~Berman, J.~P.~van der Schaar and P.~Sundell,
``A
 noncommutative M-theory five-brane,''
hep-th/0005026.}

\Title
 {\vbox{
 \baselineskip12pt
 \hbox{HUTP-01/A025}
 \hbox{hep-th/0105225}\hbox{}\hbox{}
}}
 {\vbox{
 \centerline{Mirror Symmetry and a $G_2$ Flop}
 }}
 \centerline{ Mina ${\rm Aganagic}$ and
Cumrun ${\rm Vafa}$}
 \bigskip\centerline{ Jefferson Physical Laboratory}
 \centerline{Harvard University}
\centerline{Cambridge, MA 02138, USA}
 \smallskip
 \vskip .3in \centerline{\bf Abstract}
{ By applying mirror symmetry to D-branes in a
Calabi-Yau geometry we shed light on a $G_2$ flop in M-theory
relevant for large $N$ dualities in ${\cal N}=1$
supersymmetric gauge theories. Furthermore, we derive superpotential
for M-theory on corresponding $G_2$ manifolds for all
A-D-E cases. This provides an effective method for
geometric engineering of ${\cal N}=1$ gauge theories
for which mirror symmetry gives exact information about
vacuum geometry. We also find a number of interesting dual
descriptions.}
 \smallskip
\Date{May 2001}
%\listtoc \writetoc
%\draft
\newsec{Introduction}
It was proposed in \ref\gopv{R. Gopakumar and C. Vafa,
``On the
gauge theory/geometry correspondence,'' Adv. Theor.
Math. Phys.
{\bf 3} (1999) 1415, hep-th/9811131.}\ that large $N$
$U(N)$
Chern-Simons gauge theory is dual to closed
topological strings on
the resolved conifold geometry\foot{ In the original
paper the
gauge group was taken to be $SU(N)$.  Evidence has
emerged
\ref\marva{M. Marino and C. Vafa, to appear.}\ that
this duality
is more natural for the $U(N)$ gauge group.}.  There
is now a
large body of evidence supporting this conjecture
including highly
non-trivial exact computations to all orders in $N$
\ref\ov{H.
Ooguri and C. Vafa, ``Knot invariants and topological
strings,''
Nucl. Phys. {\bf B577} (2000) 419, hep-th/9912123.
}\ref\lm{J.M.F.
Labastida and M. Marino, ``Polynomial invariants for
torus knots
and topological strings,'' Comm. Math. Phys. {\bf 217}
(2001) 423,
hep-th/0004196.}\ref\ind{ P. Ramadevi and T. Sarkar,
``On link
invariants and topological string amplitudes,''
hep-th/0009188
.}\ref\lmv{ J.M.F. Labastida, M. Marino and C. Vafa,
``Knots,
links and branes at large N,''
hep-th/0010102.}\ref\lm{ J.M.F.
Labastida and M. Marino, ``A new point of view in the
theory of
knot and link invariants,'' math.QA/0104180.}.\foot{
This duality
was also extended to the $SO(N)$ and $SP(N)$ cases in
\ref\sh{S.
Sinha and C. Vafa, ``SO and Sp Chern-Simons at large
N,''
hep-th/0012136.}.} On the other hand, this duality was
embedded in
type IIA superstrings \ref\cv{C. Vafa, ``Superstrings
and
topological strings at large N,'' hep-th/0008142.}\
where it was
interpreted as a geometric transition starting with
$N$ D6 branes
wrapped over $S^3$ of the conifold geometry, which
gives an ${\cal
N}=1$ $U(N)$ gauge theory in $d=4$, and ending on the
resolved
conifold geometry where the branes have disappeared
and been
replaced by flux.  In its mirror type IIB formulation
this is
closely related to the dualities considered in
\ref\kls{I.R.
Klebanov and M.J. Strassler, ``Supergravity and a
Confining Gauge
Theory: Duality Cascades and $\chi$SB-Resolution of
Naked
Singularities,'' JHEP 0008 (2000) 052,
hep-th/0006085.}\ref\man{J.
Maldacena and C. Nunez, ``Towards the large N limit of
pure N=1
Yang-Mills,'' Phys. Rev. Lett. {\bf 86} (2001) 588,
hep-th/0008001.}. Extensions of this duality to other
geometric
transitions has been considered in \ref\civ{F.
Cachazo, K.
Intriligator and C. Vafa, ``A large N duality via a
geometric
transition,'' hep-th/0103067.} (see also \ref\eoh{J.D.
Edelstein,
K. Oh and R. Tatar, ``Orientifold, geometric
transition and large
N duality for SO/Sp gauge theories,'' JHEP 0105 (2001)
009,
hep-th/0104037.}\ref\ohh{K. Dasgupta, K. Oh and R.
Tatar,
``Geometric transition, large N dualities and MQCD
dynamics,''
hep-th/0105066.}).

On the other hand the lift of the duality of \cv\ to
M-theory was
considered in \ref\ach{B.S. Acharya, ``On realizing
N=1 super
Yang-Mills in M theory,'' hep-th/0011089.}\ref\AMV{ M.
Atiyah, J.
Maldacena and C. Vafa, ``An M-theory flop as a large N
duality,''
hep-th/0011256.}\ where both sides of the duality
involve a smooth
$G_2$ holonomy manifold with some quotient action.
Moreover it
was argued in \AMV\ that the two geometries are
quantum
mechanically connected in a smooth way without any
singularities,
where $M2$ brane instantons would play a key role.
The aim of
this paper is to shed further light on this
transition. In
particular by employing a chain of dualities this
transition gets
related to type IIA string theory in the presence of
certain
branes. In this context mirror symmetry can be
employed, as
discussed in \ref\AV{M. Aganagic and C. Vafa, ``Mirror
symmetry,
D-branes and counting holomorphic discs,''
hep-th/0012041.}\ref\AKV{M. Aganagic, A. Klemm and C.
Vafa, ``Disk
instantons, mirror symmetry and the duality web,''
hep-th/0105045.}, to obtain exact quantum information
for
 the theory by relating it to a classical type IIB
geometry.
Not only this sheds light on the M-theory flop
involving a $G_2$
holonomy manifold, but it can also be viewed as a
program to
geometrically engineer a large class of
 ${\cal N}=1 $
gauge theories in the context of Calabi-Yau manifolds
with branes.

The organization of this paper is as follows:  In
section 2 we
recall the construction of special lagrangian A-branes
in ${\bf
C}^3$. In section 3 we review the lift of this
geometry to
M-theory \AKV . In section 4 we discuss the
application of mirror
symmetry to this geometry and rederive the basic
equation relating
the size of the resolved ${\bf P}^1$ and the volume of
$S^3$ \cv .
In section 5 we consider generalizations to $G_2$ manifolds
with A-D-E quotiont singularities.  In
section 6 we discuss some IR choices for the geometry
that do
affect the physical superpotential (in the dual
Chern-Simons
formulation this is related to the UV framing choice
of the knot),
and show that this ambiguity corresponds to the
integer in the
choice of the triple self-intersection of the Kahler
class in the
resolved conifold geometry \cv . In section 7 we
discuss a number
of dual descriptions, including 3 type IIB
descriptions, 3 type
IIA descriptions and 2 M-theory descriptions.  In one
of the dual
M-theory descriptions the whole geometry is replaced
by a single
M5 brane.

\newsec{Special Lagrangian D6 branes in $\bf{C^3}$}

In \AV \AKV\ type IIA string theory was studied where
the
background involves D6 branes wrapped on certain class
of special
lagrangian submanifolds $L$ embedded in the local
A-model
geometry. The families of lagrangians are similar to
the original
examples of Harvey and Lawson \ref\HL{ F.R. Harvey and
H.B.
Lawson, ``Calibrated Geometries,'' Acta Math. (1982)
47.}, that
were recently thoroughly studied by Joyce \ref\Joyce{
D. Joyce,
``On counting special lagrangian homology 3-spheres,''
hep-th/9907013.}. In fact, in the limit where all
sizes of the
cycles of the Calabi-Yau manifold are taken to infinity
and the
geometry is locally $\bf{C^3}$, the lagrangians $L$
are $exactly$
those of \HL \Joyce. In this section we review this
construction.
In the next section we recall \AKV\ how this geometry
in the limit
of strong type IIA string coupling is given by
M-theory in a $G_2$
holonomy geometry  which is topologically given by
$\bf{S^3 \times
R^4}$.

Consider $\bf{C^3}$ with coordinates $x_1,x_2,x_3$ and
a flat
metric. Following \AV\ consider the moment map $x_i
\rightarrow
|x_i|^2$, the image of which is $\bf{R^3_+}$. The
fiber of this
map is the torus of phases $x_i =|x_i| e^{i
\theta_i}$,
generically a $T^3$. We consider special lagrangian
submanifold
$L$ of $\bf{C^3}$ of topology $\bf{S^1 \times C}$
which is given
by the following set of equations
$$|x_1|^2 -|x_2|^2 = c_1$$
$$|x_3|^2 - |x_2|^2 = c_2$$
and $\sum_i \theta_i =0$.

The image of $L$ under the moment map is a line (Fig.
1) and the
moduli $c_i$  must be such that the line intersects the
boundary of
$\bf{R_+^3}$ along its one-dimensional edges, because
$L$ would
not have been a complete manifold otherwise. The
configuration
space of D6 brane wrapping $L$ is complexified by the
Wilson-line
around the $S^1$. There the $S^1$ that $L$ wraps
degenerates at a
point, so the classical configuration space is that of
three
copies of $\bf{C}$ meeting at a point.  These phases
are given by
the conditions:
\eqn\ph{\eqalign{{\rm Phase \ I\;\;}:\qquad &c_2=0,\;
c_1>0\cr
{\rm Phase \ II \;}:\qquad & c_1=0, \; c_2>0\cr {\rm
Phase \
III}:\qquad & c_1=c_2<0.}}
%

%%%%%%%%%%%%%%%%%%%%%%%%%%%%%%%%%%%%%%%%%
\bigskip
\centerline{\epsfxsize 3.5truein\epsfbox{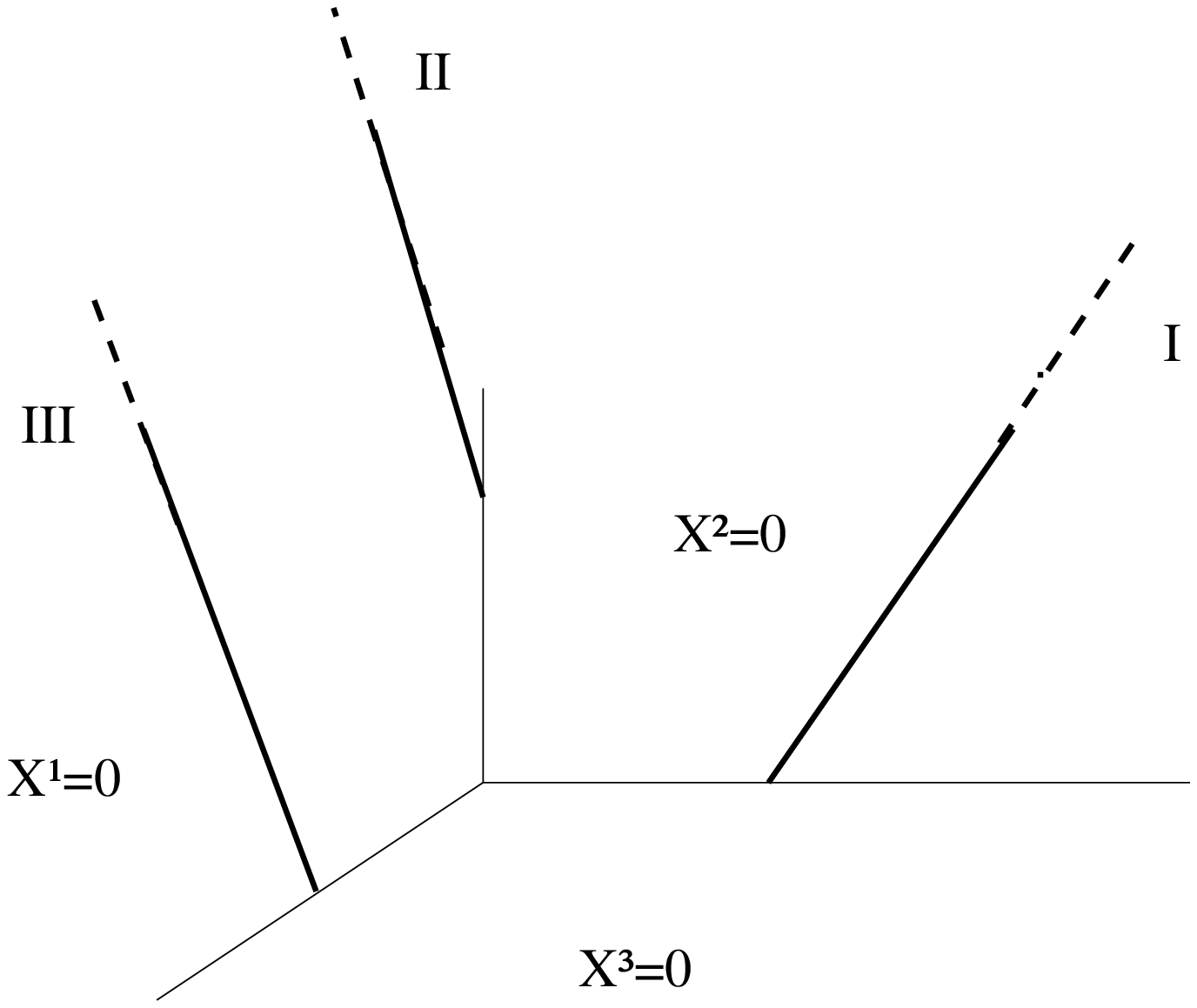}}
%\leftskip 2pc
\rightskip 2pc \noindent{\ninepoint\sl
\baselineskip=8pt {\bf
Fig.1}: The D6 brane in the three phases is depicted here.}
\bigskip
%%%%%%%%%%%%%%%%%%%%%%%%%%

It is also convenient to rewrite the equations
defining the locus
of the D6 brane as follows:  Let us restrict attention
to Phase I.
Then we can write the D6 brane locus as
\eqn\dsix{|x_1|^2-|x_3|^2=c_1, \qquad x_2={\overline
x_3}
e^{-i\theta_1}}
Note that the D6 brane has worldvolume with the
topology of ${\bf
{C\times S^1}}$ where ${\bf C}$ can be identified with
$x_3$.

There are instanton corrections to the classical
geometry which
come from worldsheets which are disks ending on the
$D6$ brane.
For example, for phase I, for a D6 brane at $c_2 = 0$,
and $c_1$ a
positive constant, the primitive disc $D$ is the
holomorphic curve
$x_2=0=x_3$ bounded by $|x_1|^2\leq c_1$. The disc
partition
function $F_{0,1}$ of the A-model topological string
computes the
exact superpotential of the type IIA string theory in
this
background. The instanton corrections can be summed up
exactly
using mirror symmetry as discussed in \AV \AKV, as
will be
reviewed below.

\newsec{Lift to M theory}

The D6 branes of IIA string theory are Kaluza-Klein
monopoles of
$M$ theory, so that the configuration we discussed
above lifts to
a purely geometric background of M-theory. In the
absence of $D6$
branes type IIA string theory on $X$ lifts to $M$
theory on
$M^7\sim X \times S^1$. Adding $D6$ branes
compactified on $L$
results in a manifold which is still locally $X\times
S^1$ but
where the $S^1$ degenerates over $L$. This makes $M^7$
into
a seven-manifold which is not a product manifold and
since the
theory has ${\cal N}=1$ supersymmetry $M^7$ must have
$G_2$
holonomy \ach\AMV\ref\gomis{J. Gomis, ``D-branes,
holonomy and
M-theory,'' hep-th/0103115.}\ref\Nuned{J.D. Edelstein
and C.
Nunez, ``D6 branes and M-theory geometrical
transitions from
gauged supergravity,'' hep-th/0103167.}.

Since the $D6$ branes are geometrized at strong
coupling the disc
$D$ discussed above must map to a complete manifold as
well. The
lift of the disc is locally $D\times S^1$ but the
$S^1$
degenerates over the boundaries of the disc, which
leads to a
manifold  of $S^3$ topology. We can model this
$S^3\subset M^7$ by
writing $|x_1|^2+|x_m|^2=c_1$ since at generic value
of $|x_1|$
this equation determines $x_m$ up to a phase which is
an $S^1$,
but at $|x_1|^2 = c_1$, the size of the $S^1$
vanishes.  The volume of $S^3$ can be identified
with $c_1$.

It is clear that there is no other non-trivial cycles
in $M^7$ and
topologically it can be identified with $\bf{S^3
\times R^4}$.  More globally we can view $x_m$ as a complex number
whose norm $|x_m|$ varies over ${\bf C}^3$ where we
identify its phase as the M-theory
circle.  Moreover $x_m$ vanishes over the special
Lagrangian submanifold $L$.
There is also a
natural M-theory projection of the enlarged moment map
from $M^7$
to a subspace of $R^3$ given by a suitable
choice of coordinates ${\tilde x}_i$ and projections ${\tilde x}_i \rightarrow
|{\tilde x}_i|^2$, for $
i=1,2,3,m$ (see Fig.2).
%%%%%%%%%%%%%%%%%%%%%%%%%%%%%%%%%%%%%%%%%
\bigskip
\centerline{\epsfxsize 5truein\epsfbox{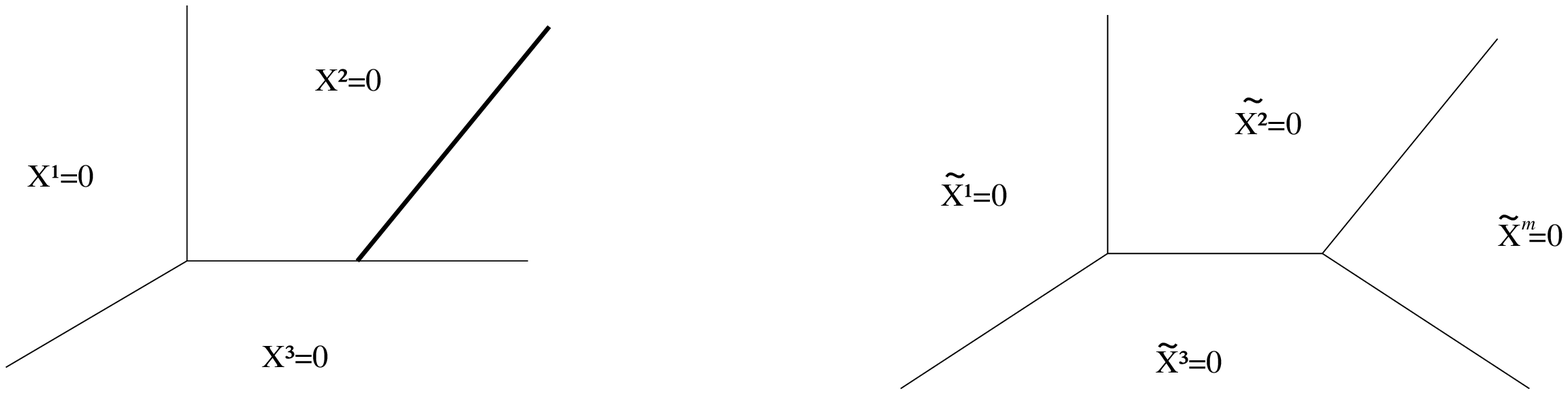}}
%\leftskip 2pc
\rightskip 2pc \noindent{\ninepoint\sl
\baselineskip=8pt {\bf
Fig.2}: The enlarged moment map from $M^7\rightarrow
R^3$ (with a suitable choice of coordinates) treats
all variables in a more symmetrical way.}
\bigskip
%%%%%%%%%%%%%%%%%%%%%%%%%%

As discussed in \AMV\ M theory has another phase  where the
volume of $S^3$ becomes negative.  Let us define $r=c_1-c_2$.
In the first phase, where $c_2=0$, $r$ is the volume of the $S^3$.
 If we take $r$ to
negative values
$r\rightarrow -r$, we can identify this as the Phase II of the
brane in the type IIA geometry.  This is where the role
of $c_1$ and $c_2$ are exchanged.  Now $c_1=0$ and $c_2>0$ denotes
the volume of the new $S^3$ (see Fig.3).

%%%%%%%%%%%%%%%%%%%%%%%%%%%%%%%%%%%%%%%%%
\bigskip
\centerline{\epsfxsize 5truein\epsfbox{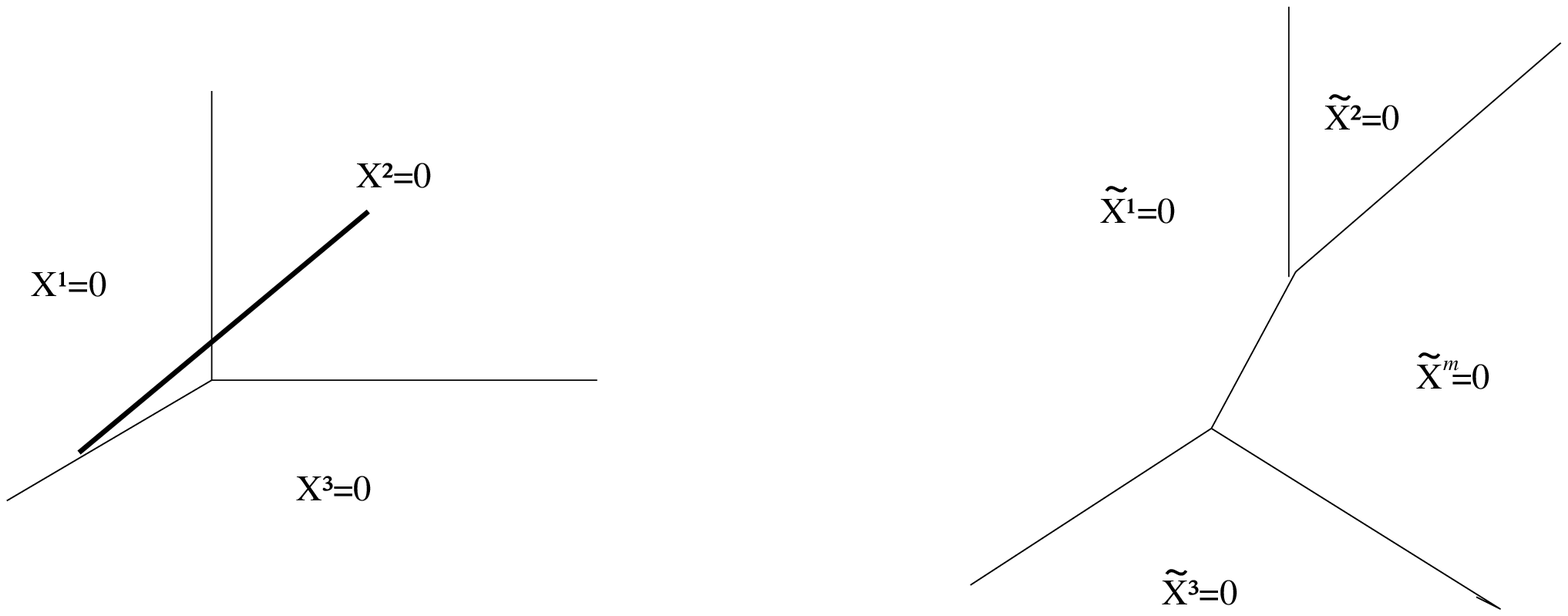}}
%\leftskip 2pc
\rightskip 2pc \noindent{\ninepoint\sl
\baselineskip=8pt {\bf
Fig.3}: The second phase and its toric base.}
\bigskip
%%%%%%%%%%%%%%%%%%%%%%%%%%

There is a third phase where $c_1 = c_2$ on the type
IIA D6 brane, i.e.  Phase III.  This corresponds
to $r=c_1-c_2=0$, and it is difficult to see this
from M-theory perspective.  However the type IIA
perspective suggests again that there is an $S^3$ that
can be blown up. The existence of the
three phase
structure in this context has been pointed out by
Atiyah and Witten \ref\aw{M. Atiyah
and E. Witten, to appear.}.

In \ach\AMV\ the $M$ theory on this same $G_2$
manifold was
considered.  The aim there was to fully geometrize the
duality of
\cv .  In this context a different circle action was
identified
with the 11-th circle where in phase I leads to the $S^3$
being the fixed point locus, which in type IIA is interpreted
as a single D6 brane wrapping $S^3$.  In phase II this
same circle action acts freely and the corresponding geometry
in the type IIA description is the resolved conifold
geometry where $S^3$ of $M$-theory maps to the $S^2$ of type IIA
dictated by the Hopf-fibration.  This induces one unit of
RR 2-form flux through $S^2$.
This transition in
type IIA
string theory is the $N=1$ case of the transitions
considered in
\cv\ as there is only a single $D6$ brane present. It
was shown in
\cv\ that the parameters of the two descriptions,
i.e., the
complexified gauge coupling $Y$ of the D6 brane
theory, $Y
=Vol(S^3) +i \int_{S^3} C$, and the complexified size
of the two
sphere $t$ are related by
\eqn\sigmaa{1-e^{-t} =e^{-Y}.}
Since the two theories exist for any value of $t$ and
$Y$, this
can be viewed as an equation for a Riemann surface
$\Sigma$ as a
hypersurface in $C^*\times C^*$ (parameterized by
$t,Y$).  We will
now show that this equation also predicts the
existence of three
classical phases.

This is an auxiliary Riemann surface in the present
language, but
notice the following. In the language of the first
IIA/M theory
duality presented above, both the volume of the $S^3$
which is
$re(Y)$ and the size of the $S^2$, the $re(t)$ become
$geometric$
parameters.  Namely in the extreme limits of the three
phases we
have discussed, we have
\eqn\phb{\eqalign{{\rm Phase \ I\;\;}: \qquad &t=0,\;
Y\gg 0 \cr
{\rm Phase \ II\;}: \qquad  & t\gg 0, \; Y=0\cr {\rm
Phase \ III}:
\qquad & t=Y\ll 0}}
Notice that these lie, approximately, on the Riemann
surface
\sigmaa . While it is true that the classical
configuration space
of $D6$ branes on $L$ resembles the geometry of the
Riemann
surface above only approximately, and in the
degenerate limit, its
quantum geometry is in fact $exactly$ that of
$\Sigma$.
Furthermore, it appears as the $classical$ geometry of
the mirror
type IIB string theory, which we now turn to.

\newsec{Mirror symmetry}

Using the recent results on mirror symmetry \ref\hv{K.
Hori and C.
Vafa, ``Mirror symmetry,'' hep-th/0002222.}\ref\hiv{K.
Hori, A.
Iqbal and C. Vafa, ``D-branes and mirror symmetry,''
hep-th/0005247.}\ one can gain further insight into
these type IIA
geometries with branes.  In particular the mirror
B-model of the
${\bf C^3}$ is given by \AKV\
$$xz = e^{-u}+e^{-v}+1.$$
Under mirror symmetry, the D6 brane wrapped on $L$
maps to a D5
brane on a holomorphic curve. The curve is given by
$x=0$, a
choice of a point on the Riemann surface \eqn\sigmab{0
= e^{-u} +
e^{-v} +1,} and fills the $z$ plane. The $B$ model
geometry
receives no quantum corrections and the theory is
exact at the
classical level. In the classical regime of the $A$
model, as
explained in \AV \AKV\ , the $A$ and the $B$ model
geometry are
related by a simple map. In phase I and for very large
$c_1$, the
$re(u)$ can be taken to be the size of the primitive
disc. The
imaginary part of $u$ comes from the Wilson line
around the $S^1$
which is finite in this phase. The $c_2$ itself is the
real part
of $v$, which vanishes in this limit, but after the
phase
transition to II and for large values of $c_2$, it is
$v$ which is
related to the disc size. More generally however, the
configuration space is the smooth Riemann surface
$\Sigma$ . From
the identification of $u,v$ as the variables measuring
the sizes
of the discs in the two limits, it is clear that we
have the map
$$Y={\hat u} \quad t ={\hat v},$$
where $\hat u=u+i\pi$ and $\hat v=v+i\pi$ denote the
flat
coordinates \AKV\ (in our discussion below we
sometimes revert
back to the $u,v$ notation for the flat coordinate
instead of
$\hat u , \hat v$).  Thus the two Riemann surfaces
\sigmaa ,
\sigmab\ are in fact canonically identified!
Furthermore, mirror
symmetry allows us to directly compute the
superpotential of the
${\cal N}=1$ theory \AV , which explains why these
Riemann
surfaces are the same.  Namely, the equation solved in
\cv\ to
yield \sigmaa\ came from minimizing the space-time
superpotential
$W$,
\eqn\eqsup{dW/dt=N{\partial ^2F\over \partial
t^2}-Y=0}
where, in the case we are considering here $N=1$, and
where $F$ is
the prepotential of the $O(-1)+O(-1)\rightarrow {\bf
P}^1$
$$F=p {t^3\over 6}+\sum_{n>0}{e^{-nt}\over n^3}$$
The integer $p$ is given by the triple
self-intersection of the
Kahler class, but it is somewhat ambiguous because of
non-compactness of the class and depends on the
choices made at
infinity.  In \cv\ $p$ was set to a particular value, but in
principle it can be
any integer.  In this section we will  set it to
zero and
return to the general case in section 6.

On the other hand as discussed in \AV\ in the B-model
the
superpotential of the D5 brane geometry is given by
$$dW/dv= u(v)-u_0$$
where $u(v)$ is found by the condition \sigmab\ and
$u_0$ is an
additional constant term which is not fixed by mirror
symmetry and
does not affect the open topological string
amplitudes. Setting
$dW/d{ v}=0$ gives ${ u}=u_0$ and ${
v}( u)$ and
thus the Riemann surface parameterizes the space of
solutions as
we change the volume of the $S^3$ in the original
formulation \cv. Not only the condition to get the minima agree, but
as one can
readily check, the superpotential $W$ of \cv\ is the
same as the
one obtained here by the methods of \AV \AKV , namely
\eqn\supppo{W=-\sum_{n>0}{e^{-nt}\over n^2}
-Yt=-\sum_{n>0}{e^{-n{\hat v}}\over n^2}-{\hat u_0}
{\hat v}}
Thus, we can view this as an alternate derivation of
the
superpotential using mirror symmetry (and a chain of
dualities!).

\newsec{Generalizations to A-D-E quotients}
In this section we study M-theory on quotients
$M^7 = S^3 \times R^4/\Gamma$,
where $\Gamma$ is a discrete A-D-E subgroup of $SU(2)$. In one
phase $\Gamma$ fixes $S^3$ and this gives rise to the corresponding
${\cal N}=1$ supersymmetric A-D-E gauge symmetry in four dimensions.
In the phase where the $S^3$ is flopped (and has formally large negative
volume with respect to the original $S^3$) the $\Gamma$ acts freely
on $S^3$ and gives the lens space $S^3/\Gamma$. We would like
to generalize the map we found using mirror symmetry between
the sizes of the $S^3$'s to these cases as well.  For A and
D there already is a prediction based on \cv\sh .  There is no formula
known for the E-series.  In this section we rederive the result
for the A case using mirror symmetry techniques.  For the D and
E series, even though one can still use mirror symmetry in principle,
we use a shortcut to obtain the map.  The idea is to use the relation
between the superpotential and the domain walls to obtain this result.
In this context we will use the results in
\ref\achva{B. Acharya and C. Vafa,``On domain walls 
of N = 1 supersymmetric Yang-Mills in four
dimensions", hep-th/0103011.}.

\subsec{The $A_{N-1}$ case}

Start with $N$ D6 branes wrapping $S^3$ as in \cv .  The
lift of this to M-theory \ach\AMV\ is given by the $Z_N$ quotient
acting on $M^7$, which fixes the $S^3$ in phase I.  This in particular means
that we have in terms of IIA variables
$$(x_1,x_2,x_3)\rightarrow (x_1, \omega x_2, \omega^{-1} x_3
)$$
where $\omega ^N=1$.  This leads
to an $A_{N-1}$ singularity in the type
IIA geometry.
However, this is not the end of the story, as we also
have a D6
brane which is the vanishing locus of the circle
action.  Note
that this is given by the same lagrangian submanifold
we started
with except orbifolded with $Z_N$:
$$L=\bf{C\times S^1}\rightarrow \bf{C/ Z_N}\times S^1.$$
We will not repeat the phase structure analysis
discussed before,
as it is already discussed in \AMV\ach\ and limit
ourselves here
to noting how the mirror type IIB geometry is
modified.

To find the mirror of this geometry with the brane, we
first have
to find the mirror of the underlying space.  This has
been done
(in more generality\foot{This can be viewed as a
degeneration of
the $A_{N-1}$ fibered geometry over ${\bf P}^1$, for
which the
mirror is given by $xz=e^{-u} +P_N(e^{-v})+\Lambda
e^{u}$, where
we take $\Lambda \rightarrow 0$ and thereby taking the
size of
${\bf P}^1$ to infinity.}) in \ref\klmvw{A. Klemm, W.
Lerche, P.
Mayr, C. Vafa and N. Warner, ``Self-dual strings and
N=2
supersymmetric field theory,'' Nucl. Phys. {\bf B477}
(1996) 746,
hep-th/9604034.}\ref\kkv{S. Katz, A. Klemm and C.
Vafa,
``Geometric engineering of quantum field theories,''
hep-th/9609239.}\ref\kmv{S. Katz, P. Mayr and C. Vafa,
``Mirror
symmetry and exact solution of 4d N=2 gauge
theories-I,'' Adv.
Theor. Math. Phys. {\bf 1} (1998) 53,
hep-th/9706110.}\  and leads
to
$$xz=e^{-u}+P_N(e^{-v})$$
where $P_N$ is a polynomial of order $N$ in $e^{-v}$.
Putting the
extra D6 brane in the geometry freezes the polynomial
$P_N$ to a
particular value of coefficients which we will now
determine.  The
mirror D5 brane is at $x=0$, filling the $z$-space.
{}From the
A-model side in the limit where the brane is in phase
I and when
the disc instanton action is very small, the classical
picture is
accurate.  In this limit we have $u\rightarrow \infty$
and the
brane is at the classical value $v\sim 0$. Moreover
there should
be a unique point for this choice, as there is a
unique brane
allowed in the A-model geometry (the ${\bf
C/Z_N}\times {\bf
S}^1)$.  This means that $P_N$ should have $N$
degenerate roots at
$v=0$, so that (up to a choice of a constant that can
be absorbed
into redefinition of variables):
$$xz=e^{-u}-(1-e^{-v})^N$$
This leads to the Riemann surface
$$x=0 , \qquad (1-e^{-v})^N=e^{-u}$$
and, using the identification $u\leftrightarrow Y$ and
$v\leftrightarrow t$, gives the equation
\eqn\nre{(1-e^{-t})^N=e^{-Y}}
in precise agreement with the relation between $t$ and
$Y$
obtained in \cv .
Note that here there is more information than just the
choice of
the Riemann surface, which parameterizes the gauge
theory moduli.
In fact, as discussed in \cv\ the duality of \gopv\
(combined with
the results of \ref\bcov{M. Bershadsky, S. Cecotti, H.
Ooguri and
C. Vafa,``Kodaira-Spencer Theory of Gravity and Exact
Results for
Quantum String Amplitudes,'' Comm. Math. Phys. {\bf
165} (1994)
311, hep-th/9309140.}) identifies
$$t\sim \langle {\rm tr} {\cal W}^2\rangle $$
where ${\cal W}$ denotes the chiral field whose bottom component
denotes the gaugino field of the gauge theory. Here, we can
$derive$ this.  To see this, note that on one hand, shifting
of the theta angle $Y \rightarrow Y +2\pi i$ across the domain
wall is accompanied by the change of the superpotential by $\Delta
W= \langle {\rm tr} {\cal W}^2\rangle$, and on the other hand
using $Y=\hat u$ and the superpotential $W =\int u dv$, we have
$\Delta W = \hat v$ is the tension of the corresponding domain
wall \AKV\ , which provides the claimed identification.
%
%In
%particular,
%the limit of decoupling of the gauge theory from the
%bulk which
%corresponds to $Y\rightarrow \infty$, leads from \nre\
%to
%
%$$tr{\cal W}^2 \sim t \sim e^{-Y/N}$$
%
%which is the expected result for the gaugino
%condensate (note that
%$Y=1/g_{YM}^2$).
%
Thus, by viewing $t=tr{\cal W}^2,\
Y=1/
g_{YM}^2$ as functions on the Riemann surface, defined
up to
shifts by $2\pi i$, they satisfy a single relation
which is the
definition of the Riemann surface.  Thus
 the parametrization
of the Riemann surface in terms of $t,Y$ carries
important
additional physical information.  Note that $Y$ is {\it also}
the tension of a domain wall \AKV\ and the Riemann
surface can be viewed as the relation between the BPS
tensions of the two types of domain walls.  These domain
walls are realized by M5 branes wrapped over $S^3$'s
on the various phases.
As shown in \ov\ the structure
of the superpotential dictates the degeneracy of the domain walls.
In particular for cases at hand, the prediction of \ov\ is
\eqn\supdom{W=\sum_{k>0}\sum_{n=1}^{\infty} d_{k} {e^{-n  kv} \over n^2}}
where $d_k$ counts the number of primitive
of domain walls in the
class given by $k v$, i.e. where the BPS tension of the domain
wall is $k v$.
For the case at hand, as discussed in \AV\AKV\ we have
$$dW/dv =u=-N{\rm log}(1-e^{-v})=\sum_n N {e^{-n v}\over n}$$
which leads to the superpotential
\eqn\supan{W=-\sum_{n=1}^{\infty} N {e^{-nv}\over n^2}}
Comparing to \supdom we see that
$|d_1|=N$ and $d_k=0$ for all $k>0$.

On the other hand, the structure
of the domain walls for M-theory on
$G_2$ holonomy manifolds of the form
$S^3\times R^4/\Gamma$ where $\Gamma $ acts as an A-D-E subgroup
of $SU(2)$ on $S^3$ and leading to lens space, was studied
in \achva 
.
The counting of the domain
walls there was carried out upon compactification of this
theory to three dimensions, as the question of counting of
the domain walls in $four$ dimensions was
found to be somewhat ambiguous, and to depend on boundary
conditions on the domain wall.
The answer in $three$
dimensions was found to have a simple structure:  Consider
for each A-D-E group $\Gamma$ the corresponding affine
Dynkin diagram.  Then there is a {\it  primitive}
domain wall associated to each node $i$ of the affine Dynkin
diagram, with BPS charge given by the Dynkin number $a_i$.
Moreover they form bound states for all possible combinations
of the primitive domain walls
consistent with assigning fermionic statistics
to the primitive ones.
The structure of the primitive domain walls follows the same
situation studied for the case of modding out conifold $T^*S^3$
by A-D-E subgroups acting on $S^3$ done in \ref\gopakva{R. Gopakumar and
C. Vafa, "Branes and fundamental groups", Adv. Theor. Math. Phys. 2,
1998, 399-411,
hep-th/9712048}\ where it was shown that
the primitive bound states are in one to one correspondence
with irreducible representations of A-D-E with charge given
by the dimension of the irreducible representation.  This
in turn is in one to one correspondence, using the McKay correspondence,
with the nodes of the affine Dynkin diagram where
the dimension of the representation is mapped to the Dynkin index.

For the A$_{N-1}$ case, the primitive
domain walls exist only in degree one, and there
is one of each for every node of the affine Dynkin diagram (as all
the Dynkin numbers are 1), i.e., $|d_1|=1$.  This agrees with the
formula we found above \supan , and so it leads to the
identification of the primitive domain walls in 3 dimensions with
the domain wall count given by the superpotential in four
dimensions.

For a general $\Gamma$, the identification of the primitive domain walls
of the 4 dimensional theory with those of the theory in three
dimensions, which correspond
to nodes of the affine Dynkin diagram and whose BPS
charge is given by the corresponding Dynkin index,
together with the relation of
the general structure of the superpotential
\supdom\ to the counting of the primitive domain walls,
suggests the superpotential is
\eqn\impor{W=-\sum_{i=0}^{r}\sum_{n>0} {e^{-n a_i v}\over n^2}}
where $r$ denotes the rank of the corresponding A-D-E.  This
 leads to
$$u=dW/dv=-\sum_i a_i {\rm log} (1-e^{-a_i v})$$
which in turn leads to
$$\prod_{i=0}^{r} (1-e^{-a_i v})^{a_i} =e^{-u}.$$
Let us write this as
\eqn\verim{\prod_{i=0}^{r}(1-x^{a_i})^{a_i}=y}
where $x=e^{-v}$ and $y=e^{-u}$.

We have already discussed the $A_{N-1}$ case.
We now present further evidence for this in D and E cases.

\subsec{The D and E cases}
The
case of $D_N$ was studied in \sh \
using the large $N$ limit of $SO(2N)$ Chern-Simons
theory, and it was shown that there are two sources
for the superpotential, after making the transition
to the blown up ${\bf P}^1$ in the type IIA orientifold
setup:  The contribution of the genus zero closed Riemann
surface with $(2N-4)$ units of RR flux to the superpotential gives
$$W_0= - (2N-4)\sum_{n>0} {e^{-2nv}\over n^2}$$
where $v=t/2$ to compare with the notation of \sh .
Moreover the $RP^2$ diagram also contributes
to the superpotential giving,
$$W_1=-2 \sum_{n>0}{{(1- (-1)^n)e^{-nv}}\over n^2}$$
One can see that
$$dW=dW_0+dW_1=-(2N-4){\rm log}(1-x^2)-2{\rm log}{(1-x)\over
(1+x)}$$
which leads to
$$y=e^{-u}=e^{-dW}={(1-x^2)^{2N-4}(1-x)^2\over (1+x)^2}=(1-x^2)^{2N-6}
(1-x)^4$$
which is in perfect agreement with \verim, when one recalls
that for $D_N$, we have four nodes with Dynkin number $1$ and
$N-3$ nodes of Dynkin number $2$.

We can also provide further evidence for \verim\ by various
other considerations:
Note that when $y \gg 0$, which corresponds
to the volume of the original $S^3$ to be negative and large,
this equation reduces to
$$x^{\sum a_i^2} =y \rightarrow x^{|\Gamma|}=y$$
where we have used the fact that the sum of the squares
of the dimension of irreducible representations add up
to the order of the group $|\Gamma|$.
This is consistent with the fact that $-v$ represents
the volume of the flopped $S^3$ and on the flopped side
$\Gamma$ acts freely and so the corresponding volume
(including the imaginary piece) is reduced by a factor
of $|\Gamma|$.  Another general check one can make
is the following:  When the volume of the original
$S^3$ is large, i.e., when $u\gg 0$ which corresponds
to $y\sim 0$ we see that $x\sim 1$, i.e., $ v\sim 0$ gives
a solution to \verim\ given by
$$\prod_{i=0}^r (a_i v)^{a_i} =e^{-u}$$
which leads to
$$v\sim e^{-u/c_2}$$
where $c_2=\sum_{i=0}^ra_i$ is the dual coxeter number
of the corresponding group.  This is consistent with the
fact that in this limit we do have a decoupled
${\cal N}=1$ gauge theory of the A-D-E type and the
gaugino condensate, identified with $v$, is expected to be given
exactly by the above formula.

Note however, that even in this limit there are more solutions
to \verim , namely by considering $x = e^{-v}$ near other $a_i$-th roots of unity.
Note that these correspond to points where the gaugino condensate
differs by a shift
${\rm tr}{\cal W}^2\rightarrow {\rm tr}{\cal W}^2+2\pi n/a_i$.
Restoring units of $M_{planck}$,
this shows that in the decoupling
limit where $M_{planck}\rightarrow \infty$ these vacua
are infinitely far away and decoupled from the A-D-E gauge theory.
Nevertheless it has been argued in \aw\
that for some of these vacua
one can find alternative
gauge theory description by considering some discrete
fluxes turned on in the $R^4/\Gamma$, that were studied in 
\ref\deBoerPX{
J.~de Boer, R.~Dijkgraaf, K.~Hori, A.~Keurentjes, J.~Morgan, D.~R.~Morrison 
and S.~Sethi, Triples, fluxes, and strings, hep-th/0103170.},
on the original side. 

Even though we have derived the \verim\ in the $A_{N-1}$ cases
using mirror symmetry and argued for its general structure
for all A-D-E, it would be interesting to derive it
directly from mirror symmetry for the D and E cases as well.

\newsec{Framing Ambiguity}
In \AKV\ it was pointed out that the choice of the IR
geometry
does modify the quantum aspects of the theory in the
presence of
branes, and in particular it modifies the
superpotential and the
Riemann surface relevant for the mirror geometry. The
ambiguity
was reflected by an integer $p$, and for example for
${\bf C}^3$
the Riemann surface was modified to
$$e^{-\hat u +p\hat v}+e^{-\hat v}-1=0$$
One would naturally ask, given the equivalence of this
Riemann
surface to the equation found in \cv\ relating $t$ to
$Y$ whether
there is also an ambiguity in there. In fact as well known the
triple intersection of the basic non-compact 4-cycle class is
ambiguous and
depends on what one fixes at infinity. This affects
the
prepotential by introducing the classical term
$$F=p {t^3\over 6}+O(e^{-t})$$
where the classical self-intersection is taken to be
the integer
$p$. If one extremizes the superpotential with this
$F$ by solving
\eqsup\ one obtains
$$(1-e^{-t})=e^{-Y+pt}$$
exactly as expected from the above ambiguity\foot{ The
generalization of this in the $A_{N-1}$ case is
$(1-e^{-t})^N=e^{-Y+Npt}$.}
 of the Riemann surface!

\newsec{Gymnastics on the Duality Web}
In this section we relate some of the considerations
in this paper
to other dual descriptions.  All said, we will end up
with three
type IIA descriptions, three type IIB descriptions,
and two
M-theory descriptions (not counting the different
phases in each
case).

Let us start with two of the type IIA descriptions:
Consider type
IIA strings on the conifold geometry  $T^*S^3$ with
$N$ D6 branes
wrapping $S^3$.  As we noted in section 5 this is dual
to type IIA
on ${\bf C}^3/{\bf Z_N}$ with one non-compact D6
brane.  Both of
these lift to the same M-theory geometry involving a
$G_2$
holonomy background with an $S^3$ at the fixed locus
of a ${\bf
Z}_N$ action, and we used this to argue their
equivalence. Acting
by mirror symmetry on the two type IIA theories
 naturally leads to two type IIB descriptions:  The
first one
leads to $N$ D5 branes wrapping $S^2$ and the second
one gives one
non-compact D5 brane in a Calabi-Yau geometry which
involves a
deformations of $A_{N-1}$ geometry fibered over another
space.  As
discussed in \AKV\ there is also yet another dual type
IIB
description involving the web of $(p,q)$ 5-branes in a
background
of ALF-like geometries.  This can be obtained by
starting from the
M-theory geometry and identifying a $T^2$ in it
(together with the
equivalence of M-theory on $T^2$ with type IIB on
$S^1$).  In this
description the circle fibration of the ALF-like
geometry is
associated to a {\it third} circle action. This
relates to the
``extended'' moment map we discussed in section 3.  It
is natural
to use a notation which treats all three circle
actions on the
same footing, in which case the vanishing loci of the
circle
actions for the case of $N$ branes are given by Fig.
4. Note that
the disappearance of the $U(N)$ symmetry in the
flopped phase is
clear, as depicted in Fig. 4; namely after the
transition the
intermediate brane appears with multiplicity one
rather than
multiplicity $N$.

%%%%%%%%%%%%%%%%%%%%%%%%%%%%%%%%%%%%%%%%%
\bigskip
\centerline{\epsfxsize 5truein\epsfbox{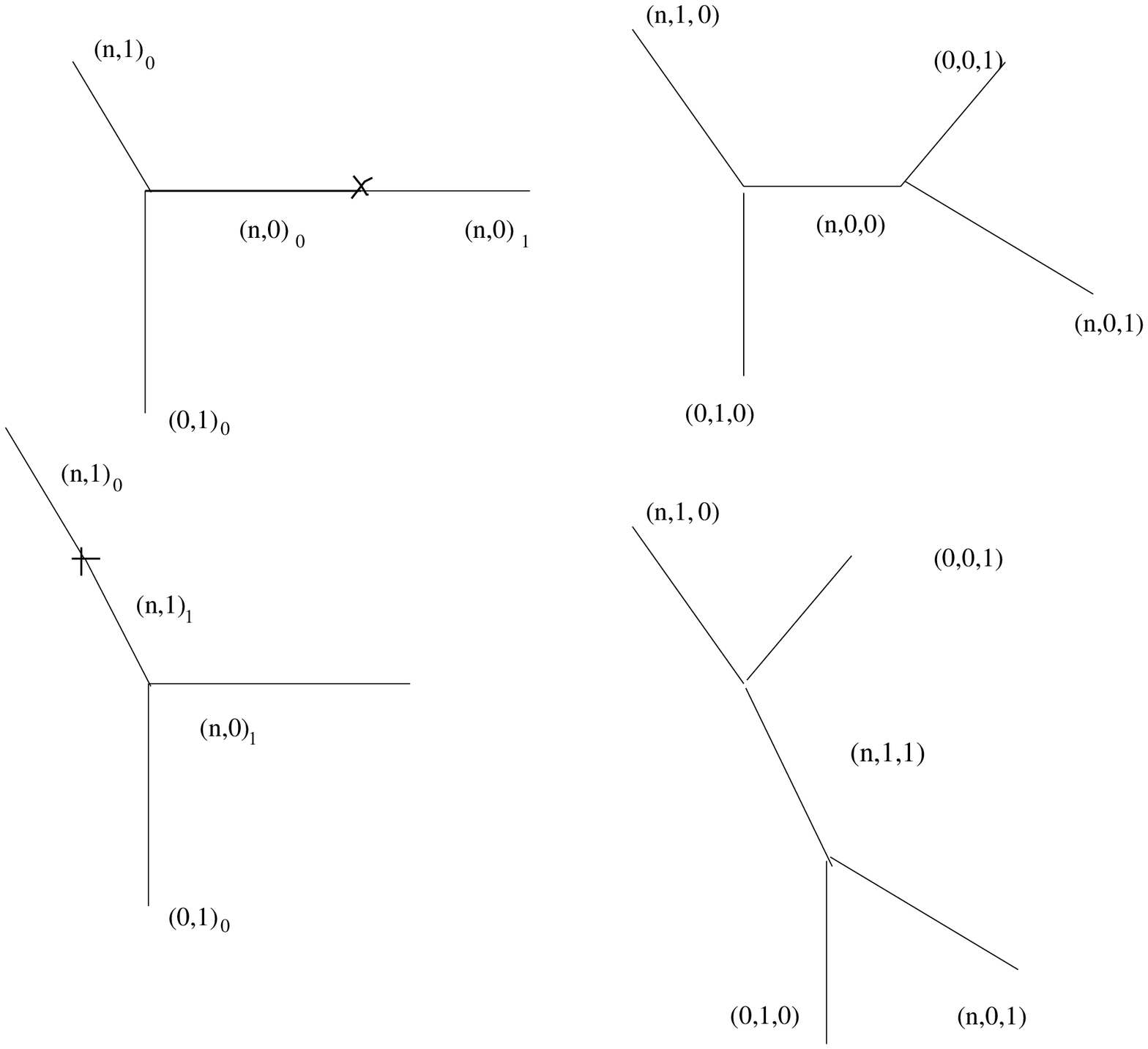}}
%\leftskip 2pc
\rightskip 2pc \noindent{\ninepoint\sl
\baselineskip=8pt {\bf
Fig.4}: Flop transition as viewed from the perspective
of $(p,q)$
5-branes of type IIB.}
\bigskip
%%%%%%%%%%%%%%%%%%%%%%%%%%

Up to now, we have three type IIB descriptions, two
type IIA
descriptions and one M-theory description.  Now we
start with the
type IIB description in the form of a local CY
geometry:
$$xz=F(u,v)$$
with a D5 brane wrapping $z$ plane at $x=0=F(u,v)$.
Let us
T-dualize this only over the circle
$$(x,z)\rightarrow (e^{i\theta} x,e^{-i\theta }z),$$
instead of the usual mirror symmetry which uses three
circles. As
discussed in \ref\oogva{H. Ooguri and C. Vafa,
``Two-dimensional
black hole and singularities of CY manifolds,'' Nucl.
Phys. {\bf
B463} (1996) 55, hep-th/9511164.}\ this gives rise to
a type IIA
description involving NS 5-brane wrapping the Riemann
surface
$\Sigma$ given by $F(u,v)=0$ and filling $R^4$.
Without the extra
D5 brane, this was used in \klmvw\ to derive the
Seiberg-Witten
vacuum geometry of $N=2$ systems, where $\Sigma $ was
identified
with the Seiberg-Witten Riemann surface.  The only
additional
ingredient here is the fact that we also have an extra
D5 brane,
filling the $z$-plane. Since we are dualizing the
circle which
corresponds to phase rotations in the z-plane, the
dual brane will
have one less dimension, i.e., it will be a
non-compact D4 brane,
ending on the NS 5-brane in the type IIA language.
This gives us
our third type IIA description.  If we now lift this
up to
M-theory, this gives a single M5 brane.  In this
formulation the
Riemann surface that we obtain is related to the brane
construction of MQCD
\ref\hoo{K.~Hori, H.~Ooguri and Y.~Oz,
``Strong coupling dynamics of four-dimensional N = 1 gauge theories from
M theory fivebrane,''
Adv.\ Theor.\ Math.\ Phys.\  {\bf 1} (1998) 1,
 hep-th/9706082}\ref\witmq{E.~Witten,``Branes and the dynamics of {QCD},''
Nucl.\ Phys.\ B {\bf 507} (1997) 658, hep-th/9706109}\ref\BIKSY{A.~Brandhuber,
 N.~Itzhaki, V.~Kaplunovsky, J.~Sonnenschein and
S.~Yankielowicz,``Comments on the M theory approach to N = 1 S{QCD}
and brane dynamics,''Phys.\ Lett.\ B {\bf 410} (1997) 27,
hep-th/9706127.}, generalizing the construction of N=2 theories in
\ref\WittenSC{E.~Witten,
``Solutions of four-dimensional field theories via M-theory,''
Nucl.\ Phys.\ B {\bf 500}, 3 (1997), hep-th/9703166.} , 
but the geometry is slightly different here.\foot{  This M5 brane
description can also be derived from the $(p,q)$ 5-brane of type IIB
along the lines discussed in \ref\kol{O. Aharony, A. Hanany and
B. Kol, ``Webs of (p,q) 5-branes, {}five dimensional field
theories and grid diagrams,'' JHEP 9801 (1998) 002, hep-th/9710116.}.}

\vglue 1cm

\centerline{\bf Acknowledgements} We would like to
thank B.
Acharya, S. Gukov, K. Hori, A. Klemm, H. Liu, M. Marino, G. Moore,
and E. Witten  for valuable
discussions.  We would also like to thank the
hospitality of the
Physics department at Rutgers university where this
work was
completed.

This work was supported in part by NSF grants
PHY-9802709 and DMS
9709694.

\listrefs
\end